\documentclass[10pt,conference]{IEEEtran}

\makeatletter
\def\ps@headings{%
\def\@oddhead{}%
\def\@evenhead{}%

\def\@oddfoot{\mbox{}\scriptsize\rightmark \hfil\thepage\hfil \leftmark\mbox{}}%
\def\@evenfoot{\mbox{}\scriptsize\rightmark \hfil\thepage\hfil \leftmark\mbox{}}}

\makeatother

\usepackage[all]{xy}
\usepackage{subfigure}
\usepackage{graphicx}
\usepackage{wrapfig}
\usepackage{url}
\usepackage{ifthen}
\usepackage{caption2}
\usepackage{setspace}
\usepackage[left=0.625in,top=0.75in,right=0.625in,bottom=1in,nohead]{geometry}

\newcommand{\commit}{\textsf{commit}}
\newcommand{\open}{\textsf{open}}

\newcommand{\mitm}{\textsf{MiTM}}

\newcommand{\dtd}{\mathsf{d2d}}
\newcommand{\dth}{\mathsf{d2h}}
\newcommand{\htd}{\mathsf{h2d}}

\newboolean{eps}
\setboolean{eps}{false}

\begin{document}
\title{\LARGE \bf Bootstrapping Key Pre-Distribution:\\
\vspace{0mm}\Large \em{Secure, Scalable and User-Friendly Initialization of Sensor Nodes}} 
\author{Nitesh Saxena and Md. Borhan Uddin\\
Computer Science and Engineering\\
Polytechnic Institute of New York University\\ 
Brooklyn, NY 11201, USA\\
{\small \texttt{nsaxena@poly.edu}, \texttt{borhan@cis.poly.edu}}
}

\maketitle

\begin{abstract}

To establish secure (point-to-point and/or broadcast) communication channels
among the nodes of a wireless sensor network is a fundamental task. To this
end, a plethora of (so-called) \textit{key pre-distribution} schemes have been
proposed in the past \cite{SPINS}\cite{last}\cite{LN03}\cite{KP03}\cite{one}.
All these schemes, however, rely on shared secret(s), which are assumed to be
somehow pre-loaded onto the sensor nodes. 

In this paper, we propose a novel method for secure initialization of sensor nodes
based on a visual out-of-band channel.  Using the proposed method, the
administrator of a sensor network can distribute keys onto the sensor nodes,
necessary to bootstrap key pre-distribution. Our secure initialization method
requires only a little extra cost, is efficient and scalable with respect to
the number of sensor nodes.  Moreover, based on a usability study that we conducted,
the method turns out to be quite user-friendly and easy to use by naive human
users.

\end{abstract}

\section{Introduction}
\label{sec:intro}
Wireless sensor nodes and sensor networks (WSN) have numerous applications in
monitoring diverse aspects of the environment.  Ready examples include
monitoring of: structural/seismic activity, wildlife habitat, air pollution,
border crossings, nuclear emission and water quality. In some applications,
sensor nodes operate in a potentially hostile environments and security measures are
needed to inhibit or detect sensor node compromise and/or tampering with
inter-node or node-to-sink communication. A large body of literature has
been accumulated in the last decade dealing with many aspects of sensor
network security, e.g., key management, secure routing and DoS detection
\cite{perrig-2004,hu-2005,last,du-2004}.

In a WSN environment, the nodes might need to
communicate sensitive data among themselves and with the sink (also
referred to as ``sink''). The communication among the nodes might be
point-to-point and/or broadcast, depending upon the application. These
communication channels are easy to eavesdrop on and to manipulate, raising the
very real threat of the so-called \textit{Man-in-the-Middle} (\mitm) attacker.
A fundamental task, therefore, is to secure these communication channels.  

\vspace{0.5mm}

\noindent \textbf{Key Pre-Distribution and the Underlying Assumption.} A number
of so-called ``key pre-distribution'' techniques to bootstrap secure
communication in a WSN have been proposed \cite{SPINS,last,LN03,KP03,one}.
However, all of them assume that, before deployment, sensor nodes are somehow
pre-installed with secret(s) shared with other sensor nodes and/or the sink.  The
TinySec architecture \cite{tiny} also assumes that the nodes are loaded with
shared keys prior to deployment. This might be a reasonable assumption in some,
but certainly not all, cases.  Consider, for example, an individual user (Bob)
who wants to install a sensor network to monitor the perimeter of his property.
He purchases a set of commodity noise-and-vibration sensor nodes at some retailer
and wants to deploy the sensor nodes with his home computer acting as the sink.
Being off-the-shelf, these sensor nodes are not sold with any built-in secrets. Some types
of sensor nodes might have a USB (or similar) connector that allows Bob to plug each
sensor node into his computer to perform secure initialization.  This would be
immune to both eavesdropping and \mitm\ attacks. However, sensor nodes might not
have any interface other than wireless, since having a special
``initialization'' interface influences the complexity and the cost of the sensor node. Also, note
that Bob would have to perform security initialization manually and separately
for each sensor node. This is not scalable since potentially many sensor nodes might be
involved. 

Also, it is important to note that keys can not always be pre-loaded during the
manufacturing phase because eventual customers might not trust the
manufacturer. Moreover, a PKI-based solution might be infeasible as it would
require a global infrastructure involving many manufacturers.\footnote{The
problem that we consider in this paper is very similar to the problem of
``device pairing'', the premise of which is also based on the fact that the
devices wanting to communicate with each other do not share any pre-shared
secrets or a common PKI with each other \cite{Balfanz02}.}

\vspace{0.5mm}

\noindent \textbf{Secure Initialization Approach.}
Therefore, the best possible strategy would be for the network administrator to
himself/herself perform the key distribution on-site.  Due to lack of hardware
interfaces (such as USB interfaces) on sensor nodes and for usability reasons,
this key distribution should be performed wirelessly. Prior key
pre-distribution schemes assume the existence of some pre-installed secret
(such as a point on a bivariate polynomial $f(x,y)$ in \cite{KP03}) using which
the shared keys can be derived.  Therefore, the task of key distribution is
reduced to establishing a secure channel between the administrator's computer
(the sink node) and each node. The resulting secure channels can in turn be
used to securely transfer, from the sink to each node, the shared secrets necessary to bootstrap key pre-distribution. Since the
administrator might need to initialize a large number of sensor nodes, the process
needs to be repeated in batches. The larger the number of sensor nodes in each
batch, the more \textit{scalable} is the secure initialization method. 

\vspace{0.5mm}

\noindent \textbf{Prior Work: Message-In-a-Bottle.} A sensor network
initialization method, called ``Message-In-a-Bottle'' (MiB), with the above
properties was recently proposed by Kuo et al. \cite{MiB07}. In MiB, the key
distribution takes place inside a Faraday Cage, which is used to shield
communication from eavesdropping and outside interference. MiB can support key
distribution onto multiple sensor nodes\footnote{Although it is not clear how many
motes at one time.} in a batch and from the administrator's perspective, it
is quite user-friendly.  However, it has some drawbacks. The first problem is
the need to obtain and carry around a specialized piece of equipment -- a
Faraday Cage.  As illustrated in \cite{MiB07}, building a truly secure Faraday
Cage is a challenge. The cost and the physical size of the Cage can be
problematic.  In other words, only a very few sensor motes could be supported in
each batch with a reasonably priced and reasonably sized cage.  The second
drawback with MiB is that if the initialization process fails for only one
sensor node or if there is an error (e.g., if the cage was not properly closed), the
entire batch of sensor nodes needs to be re-initialized and re-keyed from scratch.
Third, a batch of sensor motes must consist of homogeneous sensor motes with similar
weights (the weight is used to calculate the number of motes inside the Cage
\cite{MiB07}).  Fourth, at least one additional mote (called ``keying
device'') that possesses a physical interface, such as USB connector, is
needed. This increases both the cost and the complexity of the system.  

\vspace{0.5mm}

\noindent \textbf{Out-of-Band Channels.} To address the aforementioned
drawbacks with MiB, we consider an alternative approach based on out-of-band
(OOB) channels. The OOB (audio, visual or tactile) channels have recently been
utilized in the context of secure device pairing application
\cite{Balfanz02,MPR05,lac05,seka06}, used to establish shared keys between two
previously un-associated devices (we review these methods in Section 
\ref{sec:related}). Unlike the wireless communication channel,
the OOB channels are both perceivable and manageable by the human user(s)
operating the devices, and thus can be used to authenticate information
exchanged over the wireless channel. Unlike the wireless channel, the attacker
can not remain undetected if it interferes with the OOB channel (although it
can still eavesdrop).

\vspace{0.5mm} \noindent \textbf{Our Contributions.} Based on the protocol of Saxena et al.\
\cite{seka06}, we develop a novel initialization method using a visual OOB
channel. The underlying visual channel consists of blinking LEDs\footnote{Most
commercially available sensor motes possess multiple (typically three) LEDs. (Refer to Mica2 specifications:
\url{http://www.xbow.com/Products/Product_pdf_files/Wireless_pdf/MICA2_Datasheet.pdf)}}
as transmitters on sensor nodes and a video camera on the administrator's computer.
The design of such a channel using multiple LEDs solves an open problem posed
in \cite{seka06} and finds a useful application in the context of key
distribution for a sensor network. 

Based on our current experiments, we show that with a cheap web cam connected
to a laptop computer, we are efficiently able to use the above visual channel
to securely initialize 16 sensor nodes per batch. In addition, we perform a
usability testing of the proposed method, which shows that the method is both
user-friendly as well as robust to errors.  

As opposed to MiB \cite{MiB07}, our proposal is based upon public-key
cryptography. We note, however, that most commercial sensor motes are efficiently
able to perform public key cryptography \cite{secon}.


\section{Related Work}
\label{sec:related}

The problem of secure sensor node initialization has been considered only recently.
Prior to MiB method of \cite{MiB07} (which we reviewed in the previous
section), the following schemes were proposed. The ``Shake-them-up'' \cite{stu}
scheme suggests a simple manual technique for pairing two sensor nodes that involves
shaking and twirling them in very close proximity to each other, in order to
prevent eavesdropping.  While being shaken, two sensor nodes exchange packets and
agree on a key one bit at a time, relying on the adversary's inability to
determine the sending node.  However, it turns out that the sender can be
identified using radio fingerprinting \cite{finger} and the security of this
scheme is uncertain.  

Another two related schemes are: ``Smart-Its Friends'' \cite{sif} and ``Are You
with Me?'' \cite{youwith}. Both use human-controlled movement to establish a
secret key between two devices. In addition to having the same problems as
``Shake-Them-Up'', these schemes require an accelerometer on each sensor node to
measure movement. Most sensor nodes can not afford to have accelerometers. 

The initialization method that we propose in this paper is similar to the
device pairing schemes that use an OOB channel. Thus, we also review most
relevant device pairing methods and argue whether or not they can be extended
for the application of sensor node initialization. In their seminal work, Stajano
and Anderson \cite{SA99} proposed to establish a shared secret between two
devices using a link created through a physical contact (such as an electric
cable). As pointed out previously, this approach requires interfaces not
available on most sensor motes. Moreover, the approach would be unscalable. 

Balfanz, et al.\ \cite{Balfanz02} extended the above approach through the use
of infrared as an OOB channel -- the devices exchange their public keys over
the wireless channel followed by exchanging (at least $80$-bits long) hashes of
their respective public keys over infrared. Most sensor motes do not possess
infrared transmitters. Also, infrared is not easily perceptible by humans. 
 
Based on the protocol of Balfanz et al.\ \cite{Balfanz02}, McCune et al.
proposed  the ``Seeing-is-Believing'' (SiB) scheme \cite{MPR05}. SiB involves
establishing two unidirectional visual OOB channels -- one device encodes the
data into a two-dimensional barcode and the other device reads it using a photo
camera. To apply SiB for sensor node initialization, one would need to affix a
static barcode (during the manufacturing phase) on each sensor node, which can be
captured by a camera on the sink node. However, this will only provide
unidirectional authentication, since the sensor nodes can not afford to have a
camera each. Note that it will also not be possible to manually input on each
sensor node the hash of the public key of the sink, since most sensor nodes do not
possess keypads and even if they do, this will not scale. 

\begin{figure*}[htbp]
\centering
\framebox[6.5in]{
\begin{minipage}[ht]{6.0in}
\centering
{\footnotesize{
\begin{tabular}{ccc}

$\underline{{\textbf A~\mbox{(sensor node)}}}$ & & $\underline{{\textbf B~\mbox{(sink)}}}$\\
\\
Pick $R_A \in \{0,1\}^k$ & & \\
$(c_A,d_A) \leftarrow \commit(pk_A, R_A)$ ~~~~~~~~&$\xymatrix@1@=80pt{\ar[r]^{pk_A,c_A}&}$ &Pick $R_B \in \{0,1\}^k$ \\
& $\xymatrix@1@=80pt{& \ar[l]_{pk_B,R_B}}$ & \\
 
& $\xymatrix@1@=80pt{\ar[r]^{d_A}&}$ & \\

$SAS_A = R_B \oplus H_{R_A} (pk_B)$ &$\xymatrix@1@=80pt{\ar@{=>}[r]^{SAS_A}&}$ &\\

& & $R_A \leftarrow \open(pk_A,c_A, d_A)$ \\
& $\xymatrix@1@=35pt{& \ar@{==>}[l]_{b}}$$\xymatrix@1@=35pt{& \ar@{-->}[l]_{b}}$& $b \leftarrow (SAS_A == R_B \oplus H_{R_A} (pk_B))$\\

Accept $pk_B$ as $B$'s public key if & & Accept $pk_A$ as $A$'s public key if \\
{$b=1$} & & {$b=1$}\\

\hline
\end{tabular}
\begin{tabular}{l}
 $\xymatrix@1@=30pt{& \ar@{<->}[l]_{}}$: the wireless channel\\
 $\xymatrix@1@=30pt{& \ar@{<=}[l]_{}}$: the unidirectional $\dtd$ channel\\
 $\xymatrix@1@=30pt{& \ar@{-->}[l]_{}}$: the $\dth$ channel\\
 $\xymatrix@1@=30pt{& \ar@{==>}[l]_{}}$: the $\htd$ channel\\
 $pk_A, pk_B$: public keys of devices $A$ and $B$\\
 \commit() and \open(): functions of a commitment scheme based on random oracle model (in practice, SHA-1/MD5)\\
 $H()$: hash function drawn from an almost universal hash function family\\
\end{tabular}
}}
\end{minipage}

}
\caption{The protocol by Saxena et al. \cite{seka06} based on the SAS protocol of \cite{PV06}}
\label{fig:seka}
\end{figure*}
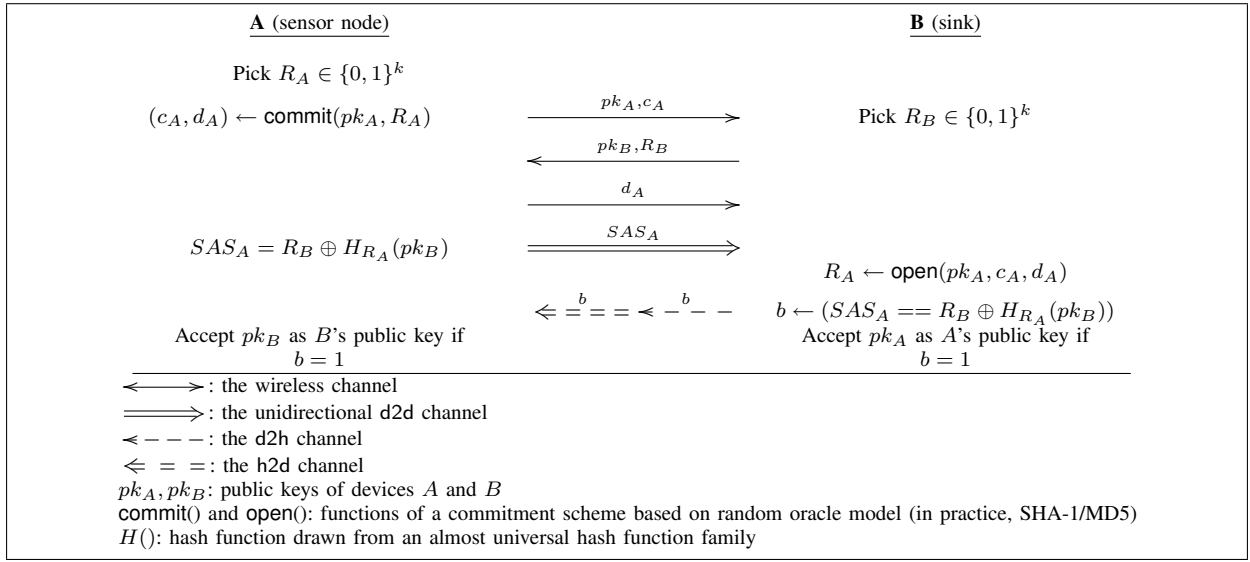

Saxena et al. \cite{seka06} proposed a new scheme based on visual OOB channel.
The scheme uses one of the protocols based on Short Authenticated Strings (SAS)
\cite{PV06}, \cite{Nyberg05}, and is aimed at pairing two devices (such as a
cell phone and an access point), only one of which has a relevant receiver
(such as a camera).  The protocol is depicted in Figure \ref{fig:seka} and as
we will see in the next section, this is the protocol that we utilize in our
proposal. In this paper, we extend the above scheme to a ``many-to-one''
setting applicable to key distribution in sensor networks. Basically, the novel
OOB channel that we build consists of multiple devices blinking their SAS data
simultaneously, which is captured using a camera connected to the
sink.

Recently, Soriente et al. \cite{hapadep} consider the problem of pairing two
devices based on an audio channel. Their scheme can be based on the protocol of
\cite{seka06}, with the unidirectional SAS channel consisting of one device
encoding its SAS data into audio, and the other device capturing it using a
microphone. Extending this scheme to initialize multiple sensor nodes in a scalable manner seems
hard as it will be hard to decode simultaneously ``beeping'' nodes.

There are a variety of other pairing schemes, based on manual
comparison/transfer of OOB data: \cite{lac05,uka06} can not be used on sensor nodes
as they require displays; \cite{beda,sr07} are applicable on sensor nodes but would
not scale well due to their manual nature.

\section{Communication and Security Model, and the Underlying Protocol}
\label{sec:protocols}

\noindent \textbf{Model.}
The protocol that we utilize in our initialization method is based upon the
following communication and adversarial model \cite{Vaudenay05}. The devices
being paired are connected via two types of channels: (1) a short-range,
high-bandwidth bidirectional wireless channel, and (2) auxiliary low-bandwidth
physical OOB channel(s). Based on device types, the OOB channel(s) can be
device-to-device ($\dtd$), device-to-human ($\dth$) and/or human-to-device
($\htd$). An adversary attacking the pairing protocol is assumed to have full
control on the wireless channel, namely, it can eavesdrop, delay, drop, replay
and modify messages. On the OOB channel, the adversary can eavesdrop on but can
not modify messages. In other words, the OOB channel is assumed to be an
authenticated channel. The security notion for a pairing protocol in this
setting is adopted from the model of authenticated key agreement due to Canneti
and Krawczyk \cite{CK01}. In this model, a multi-party setting is considered
wherein a number of parties simultaneously run multiple/parallel instances of
pairing protocols. In practice, however, it is reasonable to assume only
two-parties running only a few serial/parallel instances of the pairing
protocol. For example, during authentication for an ATM transaction, there are
only two parties, namely the ATM machine and a user, restricted to only three
authentication attempts. The security model does not consider denial-of-service
(DoS) attacks. Note that on wireless channels, explicit attempts to prevent DoS
attacks might not be useful because an adversary can simply launch an attack by
jamming the wireless signal.

In a communication setting involving two users restricted to running three
instances of the protocol, the SAS protocol of \cite{seka06} need to transmit only $k$
($=15$) bits of data over the OOB channels. As long as the cryptographic
primitives used in the protocols are secure, an adversary attacking these
protocols can not win with a probability significantly higher than $2^{-k}$
($=2^{-15}$). This gives us security equivalent to the security provided by
5-digit PIN-based ATM authentication.


\noindent \textbf{Protocol.}
The protocol that we utilize \cite{seka06} is depicted in Figure 1
(we base the protocol upon the SAS protocol of \cite{PV06}, although it can
similarly work with other SAS protocol \cite{Nyberg05} as well). The protocol works as
follows.  Over the wireless channel, devices A (sensor node) and B (sink) follow the
underlying SAS protocol. Then a unidirectional OOB channel is established by
device A transmitting the SAS data, over the $\dtd$ channel. This is followed by device B comparing the
received data with its own copy of the SAS data, and transmitting the resulting
bit b of comparison over the $1$-bit $\dth$ OOB channel (say, displayed on its screen). Finally,
the user reads the transmitted bit $b$ and accordingly indicates the result to
device A by transmitting the same bit $b$ over an $\htd$ input channel.

For our application of secure initialization of sensor nodes, we execute the
protocol of \cite{seka06} in a ``many-to-one'' setting. Basically, the sink
runs serial or (preferably) parallel instances of the pairing protocol over the
wireless channel with each of the $n$ sensor nodes belonging to a batch.  The SAS
data, however, is transmitted simultaneously from each sensor node to the sink.
Since the SAS data is transmitted simultaneously by each sensor node, the sink has
no efficient way to figure out what SAS value was transmitted by which of the
sensor nodes it discovered over the wireless channel. Therefore, the sink accepts
the key distribution on a particular sensor node A if the SAS value (derived from
information transmitted over the wireless channel) corresponding to A matches
with any of the $n$ SAS values received over the SAS channel.  Sensor node A is
therefore accepted with a probability at most $n2^{-k}$ instead of $2^{-k}$ as
in the original ``one-to-one'' setting. One can show that such a
``many-to-one'' variant can be proven secure. In other words, one can show that
if there exists an adversary who breaks, with a probability significantly
better than $n2^{-k}$, the above many-to-one variant of the protocol of
\cite{seka06}, then there exists another adversary who can break the protocol
of \cite{seka06} with a probability significantly better than $2^{-k}$.  We
omit this proof in this paper and concentrate on the design and implementation
of the underlying SAS channel, using multiple LEDs and a video camera. Note
that in order to achieve the same level of security offered by a 5-digit
PIN-based authentication (as mentioned above), the length of the SAS data
should now be $15 + log_2(n)$. 

The security of our initialization method is equivalent to the security of the
underlying SAS protocol, under the assumption that the administrator correctly
discards the sensor nodes based on the result (bit $b$ corresponding to each sensor node)
indicated by sink.


\section{Our Proposal: Secure Initialization using a Visual Channel} 
\label{sec:experiment}

In this section, we describe the design and implementation of an efficient,
scalable, user friendly and commercially viable method of secure initialization
for sensor nodes. The core of our solution relies on the protocol of \cite{seka06}
executed in a many-to-one setting, as mentioned in the previous section.
For transmitting the SAS data of all sensor nodes simultaneously over the visual
channel, the LEDs of sensor nodes are used for ON-OFF encoding,  and for receiving
the data, video frame based image processing is used on the receiver side. 

\subsection{Set-up of the Mechanism}

In our setup (Figure \ref{fig:all}), the administrator's computer (the
sink) is connected (using a USB interface) with a sensor node having the
functionality of a base station.  The sink is also connected with a video camera
(a web cam).  The sensor nodes and the sink communicate over the wireless channel. 
Sensor nodes have their on board displays implemented using two types of LEDs -- one
Sync LED (used for synchronizing the data transmission between the sensor node and
the sink) and at least one Data LED (used for transmitting SAS data).  The Data
LEDs can be of any color (same or different), but their color(s) should be
different from the color of the Sync LED. The blinking LEDs on sensor nodes are used
to transmit the SAS data, which is captured using the camera on the sink.  The
sink matches the received SAS data with its own copy of the acquired SAS data
for each sensor node and based on this, learns whether a particular sensor node
``passed'' or ``failed'' during the process. The sink also displays on its
screen the result corresponding to each sensor node. Based on the result indicated,
the administrator must remove or turn off the failed sensor nodes.  In case the sink
is also connected with a printer, the screen indicating the result can also be
printed, to better assist the administrator.

\subsection{Role of the Administrator}

The administrator needs to follow the steps shown in Figure \ref{fig:role2}.
On completion of Step 5, the sink makes use of the resulting secure channels between
itself and each sensor node to bootstrap any of the key pre-distribution schemes,
e.g., \cite{KP03}.

\begin{figure}[h]

\framebox[3.5in]{
\begin{minipage}[ht]{3.4in}
{\footnotesize{
\vspace{1mm}

\textit{Step 1.} The administrator turns on the sensor nodes and places them on a table, one by one.
\vspace{1mm}

\textit{Step 2.} The administrator presses the ``Start'' button on the sink. This triggers
the sink to sense the nearby sensor nodes and signal them over the wireless channel
to start an instance each of the protocol of Figure \ref{fig:seka}. Once done with their SAS data
computation, the sensor nodes show a ``Ready'' signal to the administrator 
by lighting up their red LEDs, and the sink shows
the message ``Focus the Camera on Ready Sensor nodes and Press OK''.
\vspace{1mm}

\textit{Step 3.} The administrator adjusts the camera accordingly to capture the LEDs of
the ready sensor nodes and presses the ``OK'' button on the sink. The sink sends a
``Start Transmission'' signal over wireless channel to all sensor nodes
simultaneously to transmit their SAS data. All the sensor nodes transmit their SAS data
simultaneously and the camera on the sink captures and decodes the data, 
and shows the result on the screen and/or prints it out. 
\vspace{1mm}

\textit{Step 4.} The administrator turns off the failed sensor nodes based on the
on-screen or printed output. The turning off of a sensor node is to be implemented
in such a manner that it is equivalent to the sensor node \textit{rejecting} the
protocol instance it executed with the sink. If the administrator does not turn
off a particular sensor node, within an (experimentally determined) time period
$\Delta$, by default, the protocol instance will be accepted by the
sensor node. (The default acceptance mechanism is adopted in order to improve
the usability of our method. Under normal circumstances, i.e., when no attacks
or errors occur, the administrator does not need to turn off any sensor node.)

\vspace{1mm}

\textit{Step 5.} Steps 1-4 are repeated, batch by batch, until all sensor nodes 
successfully initialized.

\vspace{1mm}
}}
\end{minipage}
}
\caption{The Administrator's Role}
\label{fig:role2}
\end{figure}

\subsection{Design and Implementation}
Our sensor node initialization method requires three phases: (1) the device
discovery phase, whereby the sink discovers each sensor node (over the wireless
channel)\footnote{The sink as well as the sensor nodes need to know the actual
number $n$ of sensor nodes being initialized in one batch, since the length $k$ of
random nonces $R_A$ and $R_B$ and of $SAS_A$ in the protocol of Figure
\ref{fig:seka}, should ideally be equal to $15+log_2(n)$ (as discussed in
Section \ref{sec:protocols}).  However, an adversary might influence the value
of $n$ the sink and the sensor nodes determine by sensing over the wireless channel.
Therefore, one can hard-code the value of $k$ on the sensor nodes and on the sink,
based on the expected maximum number of sensor nodes to be initialized in a batch.
For example, one can safely set $k$ to be 20, if it is expected that at most
only 32 sensor nodes will be initialized in a batch.}, (2) protocol execution phase,
whereby the first three rounds of the SAS protocol of Figure \ref{fig:seka} are
executed between the sink and each sensor node, and (3) the SAS data transmission,
whereby the sensor nodes simultaneously transmit their SAS data, the sink captures
them, matches each of them with the local copies and accordingly indicates to the
administrator to discard any failed sensor nodes.

\ifthenelse{\boolean{eps}}{}
{
\begin{figure*}[htbp]
\begin{center}
\begin{minipage}[t]{0.35\linewidth}
\centering
\fbox{\includegraphics[height=2.5in,width=2.8in]{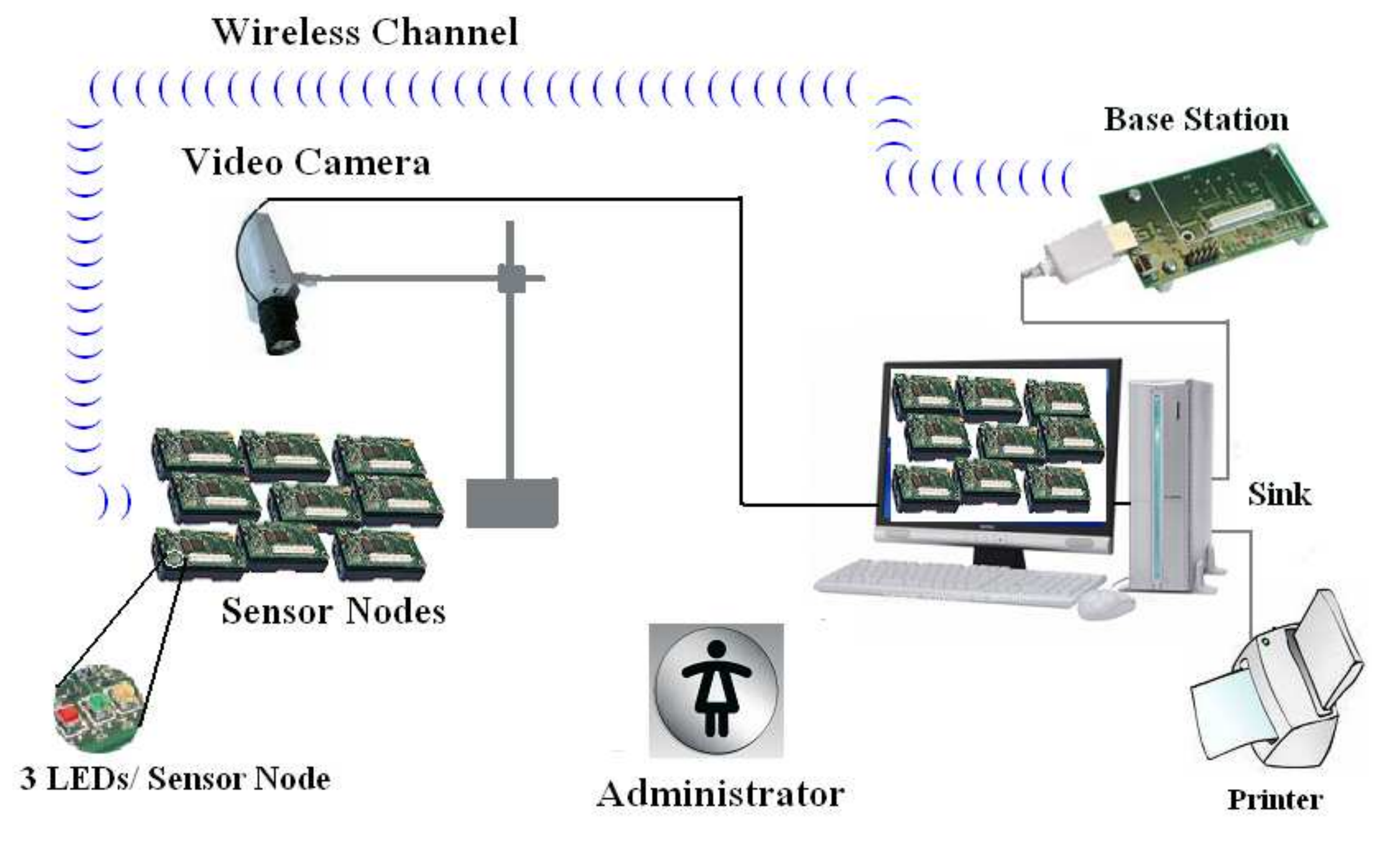}}
\caption{The Overall Set-up of Mechanism} \label{fig:all}
\end{minipage}
\hspace{0.1in}
\begin{minipage}[t]{0.62\linewidth}
\centering
\fbox{\includegraphics[height=2.5in,width=3.8in]{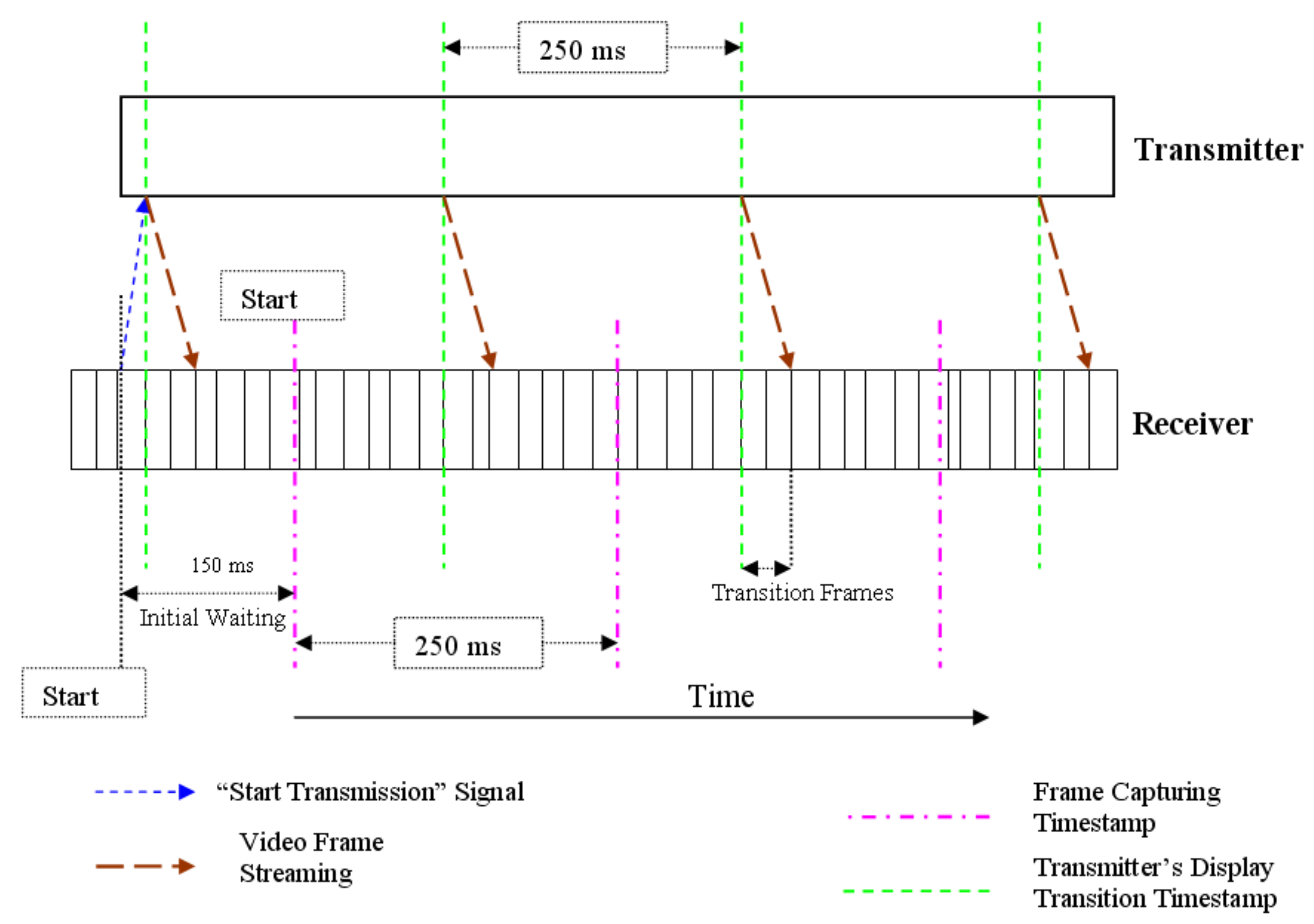}}
\caption{Synchronization of Transmission (using LEDs) and Reception (on sink) of Data}\label{fig:sync}
\end{minipage}
\end{center}
\end{figure*}

}

We were most interested in the third phase as it is an essential element of our
proposal.  To this end, we have developed an application in Microsoft Visual
C\# that simulates our sensor node initialization process. The application has two
parts -- the transmitter simulating the sensor nodes and the receiver simulating the
sink; running on two different computers.  The transmitter encodes and
transmits the SAS data using the display consisting of three blinking LEDs per
sensor node. All sensor nodes show the $i^{th}$ bit of their respective SAS data
simultaneously. The sink captures the transmitted data as a video stream using
its camera, extracts the SAS data for each sensor node, compares it with its own
local copy for the corresponding sensor node and displays the result on screen
and/or prints it out through a printer connected to it. Instead of dealing with
real sensor motes\footnote{Clearly, since we wanted to deal with a number of
sensor motes, a testbed consisting of real sensor motes was not affordable, nor was it
necessary.}, we simulated the display of sensor motes using LEDs on a breadboard,
integrated with the transmitter through the parallel port of the
transmitting computer. 

\vspace{1.5mm}

\subsubsection{Encoding using LEDs}

In our simulation, each sensor node is equipped with one Sync LED(red color LED)
for synchronization at the beginning and end of SAS data transmission and 
two Data LEDs(green color LEDs) for transmitting the SAS data.
We simulated the display of a total of 16 sensor nodes on a breadboard(Figure \ref{fig:real}) each having
three LEDs as most commercially available sensor motes; however, our implementation 
supports an arbitrary number of LEDs (with an arbitrary physical topology) and two
distinct but not fixed color LEDs(for differentiating Sync and Data LEDs).


The sync LED (kept ``ON'' at the beginning and end of SAS data transmission; ``OFF'', otherwise) is used
 to indicate the beginning and end of the SAS data
transmission and to detect any synchronization delays,
adversarial or otherwise, between the sensor nodes and the sink.
 
The data LEDs are used for SAS data transmission by indicating different bits
(`0'/`1') using different states (OFF/ON) of LEDs. If N is the number of Data
LEDs, the transmitter can display N bits of SAS data at a time. The states of the
sync and data LEDs are kept unchanged for a certain time period (named ``hold
time''; experimentally determined as 250ms); so that, a stable state (named
``BitFrame'') can be easily captured in the video stream of the receiver video
camera. After every 250 ms, next N bits of the SAS data are simultaneously shown by each sensor node in the next
frame. This process continues until all bits of SAS data are transmitted. If the
last frame does not have N number of SAS bits to show, the beginning required
LEDs show the data bits and the remaining are kept OFF.

For discovering the location, color, dimension of LEDs for each sensor node at
the receiver side, two extra frames are needed at the beginning of data transmission
 -- an ``All-ON'' frame having all
LEDs in ON state and an ``All-OFF'' frame having all LEDs in OFF state.
In addition to All-ON and All-OFF frames, another frame is required at the end of SAS data transmission, to
detect synchronization delays having the Sync LED in ON state and the data
LEDs in OFF state. Therefore, overall a total of three extra
frames are required. Thus, for 20-bit SAS data transmission(recall that $[15+log_2(16)]$-bit long SAS
is required for 16 sensor nodes) the total number of frames to be transmitted is
$\lceil\frac{20}{N}\rceil+3$, which yields a total transmission time of
$(\lceil\frac{20}{N}\rceil+3)\times 250$ ms. For transmitting 20-bit SAS data using N=2
data LEDs, there is requirement of a total of 13 frames and thus a total of 3.25 seconds of
transmission and capturing time.



\vspace{1.5mm}

\noindent \subsubsection{Decoding using a Video Camera}

For successfully decoding the data transmitted using the LEDs of sensor nodes, the
receiver video camera must have a frame rate higher than the transmission rate.
If frames are not carefully captured from the video stream, there is a
likelihood of obtaining the counterfeit frames, which contain the transition
state of LEDs.  


\noindent \textbf{Resolving the Timing Issue of Frame Capturing.} We assume
that the transmission delay of ``Start Transmission'' (ST) signal from the
receiver to the transmitter is negligible (5-6 ms) compared to the ``hold
time'' (HT) (of 250 ms) and the receiver video camera also has a delay (about
30-40 ms, since most common cameras have a rate of 30-40 frames per second) of
capturing the frame from video stream. Bases on this assumption, the receiver
captures the first frame from the video stream after a time, equal to
0.6$\times$ HT (i.e. after 150 ms), termed as ``initial waiting'' (IW), after
sending the signal.  The sink pre-calculates capturing (saving frames into
memory from video stream buffer) timestamps for all frames by adding the IW +
(HT (250ms) $\times$``frame\_index''), with the timestamp of sending of the ST
signal.  The frames are captured into memory at the corresponding timestamps.
 Figure \ref{fig:sync} depicts the synchronization of transmission and
reception of SAS data. In this figure each small rectangle on the receiving
window denotes a video frame of video stream and brown arrow marked with
``Video Frame Streaming'' denotes the propagation of transmitted signal to
streamed frame in the video stream, which implies that there is some
propagation delay of an input transition from transmitter's side to the receiver's
video stream.

\noindent \textbf{Detection of LEDs and Retrieval of SAS data.}
The frames are processed after the completion of capturing of all required 
frames. 
Our LED location and dimension detection algorithm is simple yet fast,
robust and efficient, unlike existing object/face detection algorithms
\cite{ImgProcRef01,ImgProcRef02,ImgProcRef03}.
The algorithm detects the position and dimension of LEDs
deterministically. It is able to detect any shape/geometry of LEDs unlike
\cite{ImgProcRef03} and does not require any
prior training unlike \cite{ImgProcRef01,ImgProcRef02}. The algorithm uses
the color threshold adjustment technique like \cite{ImgProcRef06} to
detect the position and dimension of LEDs.

The maximal differences of RGB values, $max(dR,dG,dB)$ (denoted as $\mu$), of each pixel of
All-OFF and All-ON frames are measured and kept in memory. Using 
a threshold value for $\mu$, bit-strings are built for each row of
pixels. For example, if $\mu$ exceeds a certain threshold, the
corresponding bit in the string becomes `1', otherwise it becomes a `0'.

Each bit-string is matched against a regular expression for consecutive '1's.
For each matching bit-string, its center is calculated and its safeness and
centeredness as an LED center is checked by matching against the already
explored LEDs and exploring only the nearby pixels of this center in the frame.
If its safeness and centeredness is proved, it is accepted as an LED and its
coordinates are included in the explored list of LEDs. 
This process continues up to a number of times by adjusting the threshold value
of $\mu$ and constructing the new bit strings until all LEDs are detected. In
Figure \ref{fig:det-led}, we show an example of detection of LEDs from the
bit-string.

\ifthenelse{\boolean{eps}}{}
{
\begin{figure*}[htbp]
\begin{center}
\begin{minipage}[t]{0.30\linewidth}
\centering
\fbox{\includegraphics[height=1.3in,width=2in]{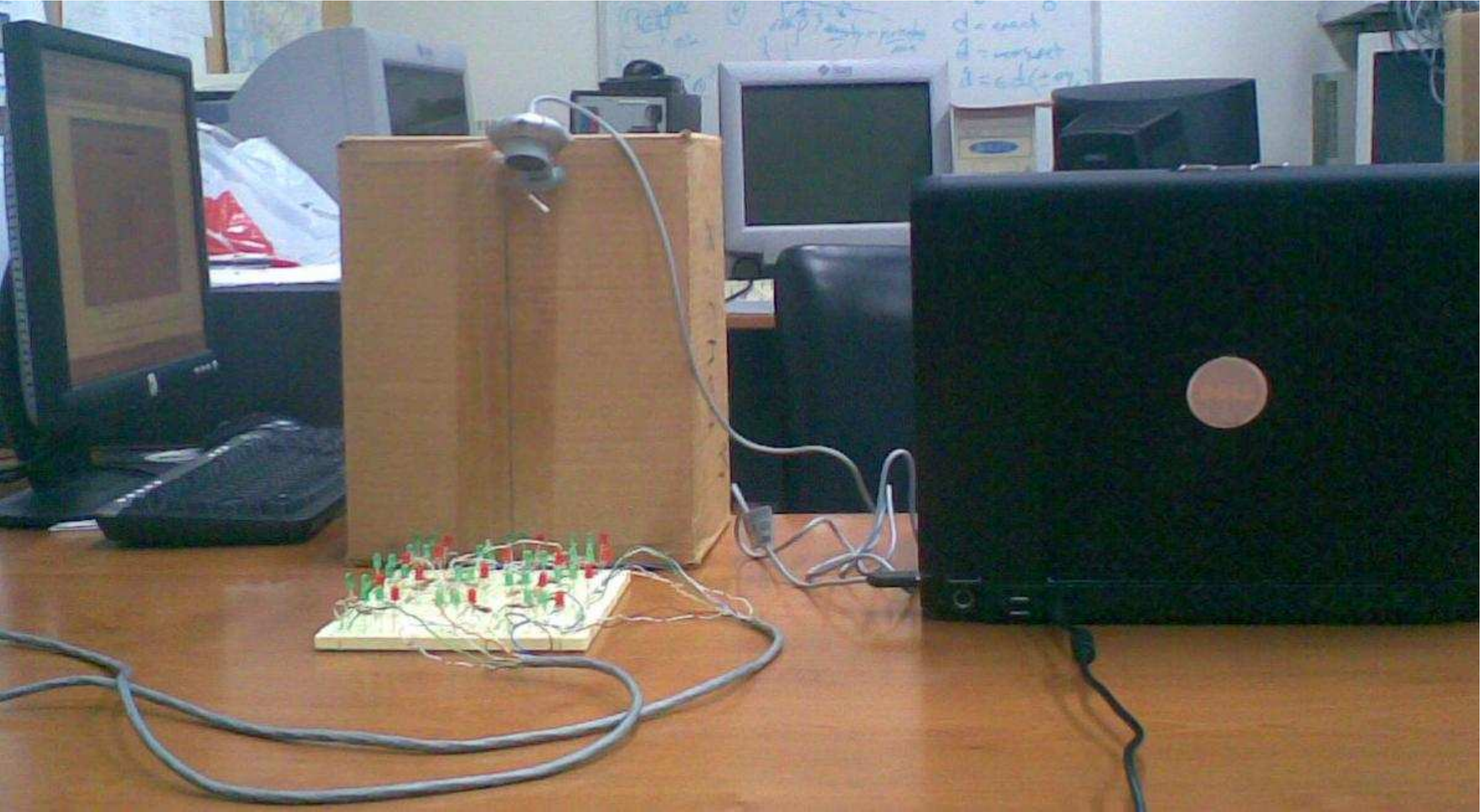}}
\caption{Setup: Receiver is Web camera, Transmitters are LEDs on Breadboard} \label{fig:sim}
\end{minipage}
\hspace{0.1in}
\begin{minipage}[t]{0.30\linewidth}
\centering
\fbox{\includegraphics[height=1.3in,width=2in]{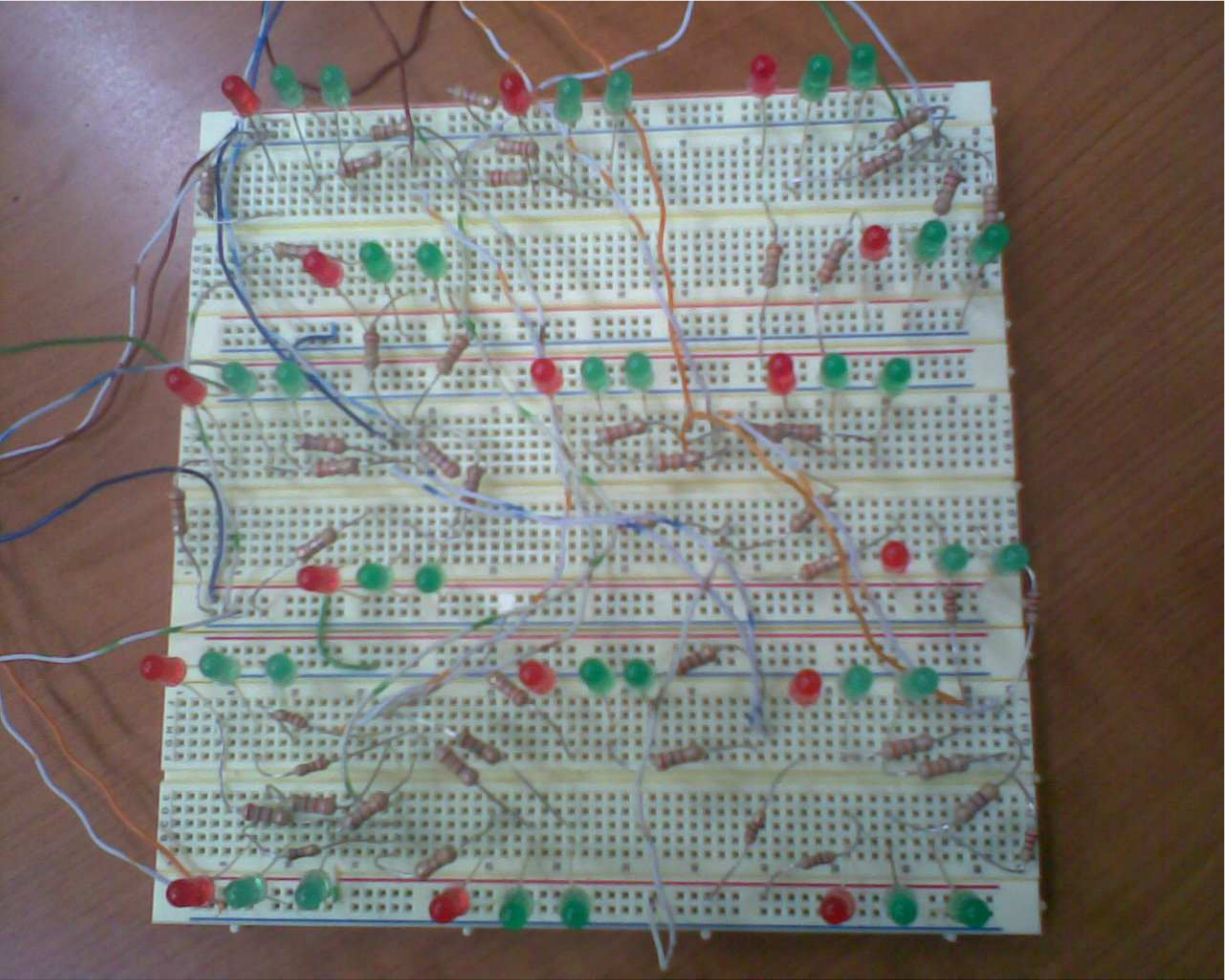}}
\caption{Transmitter: Breadboard with 48 LEDs Simulating 16 Sensor Nodes} \label{fig:real}
\end{minipage}
\hspace{0.1in}
\begin{minipage}[t]{0.30\linewidth}
\centering
\fbox{\includegraphics[height=1.3in,width=2in]{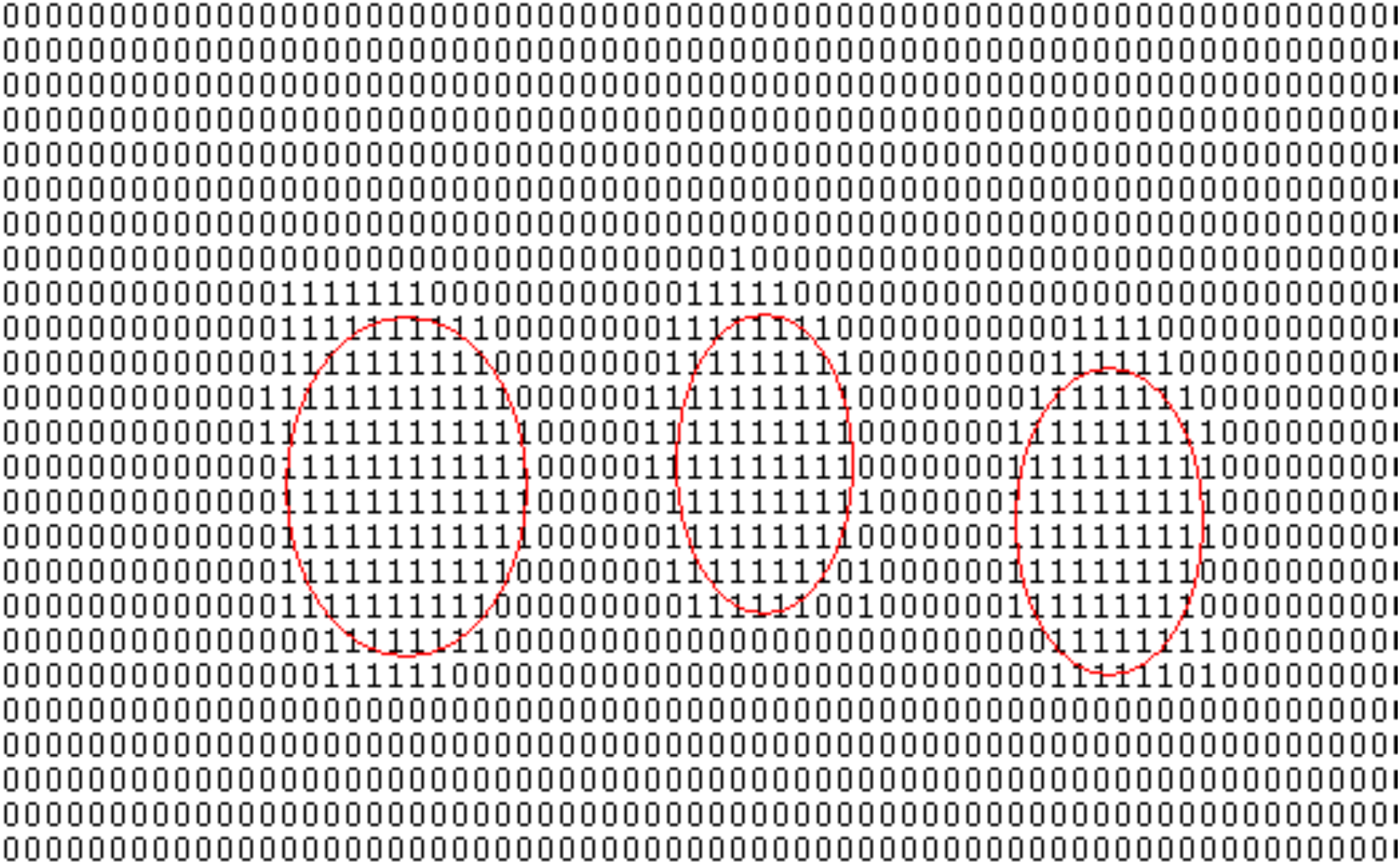}}
\caption{Detected LEDs from BitString} \label{fig:det-led}
\end{minipage}%

\end{center}
\end{figure*}

}


After successful discovery of LEDs, the length, width, average RGB values of ON
and OFF states of LED area, for each LED are stored in memory for detecting the
ON/OFF state of LEDs in subsequent BitFrames. Successfully discovered LEDs are 
clustered according to a threshold value of proximity among themselves, for identifying
the displays of different sensor nodes.


After successful detection of all sensor nodes, the data LEDs of each sensor node are
sorted according to the left-to-right and top-to-bottom ordering of
coordinates. Now SAS data for each sensor node is extracted from the BitFrames by
comparing the average RGB values of LEDs with previously saved (from All-OFF
and All-ON frames) OFF and ON state RGB values of LEDs. For each extracted SAS, the sink matches it with its
own computed list of ``free'' SAS values. If there is a match, the sink
marks the corresponding computed SAS as ``used'' and the sensor node as ``SAS
Matched''. If extracted SAS of a sensor node does not match with any free SAS values, the
corresponding sensor node and all sensor nodes having the same SAS are
marked as ``SAS Mismatched''.
Each BitFrame is then examined: the Sync LEDs of all sensor nodes
should be in the OFF state, except for the last frame, where the Sync LED should
be in the ON state and all data LEDs of all sensor nodes should be in the OFF state.
If this is not the case, it implies that a synchronization error
occurred.

If for a sensor node, both ``SAS Matched'' and ``Sync
Matched'' are true, the sink accepts the sensor node as a ``passed'';
 otherwise, it rejects the sensor node as a ``failed'' due to mismatch
of SAS and/or synchronization errors.  The LEDs of a passed sensor node are marked
with a rectangle of green color; and the LEDs of a failed sensor node are crossed
out with red color (Figure \ref{fig:res2}). Additionally, an automatic printing 
of the result-screen is done by the printer connected to the sink. By observing 
the graphical result on screen of the sink and/or the printed result, the 
administrator discards the failed sensor nodes.


\section{Experiments and Results}
\label{sec:test}
\subsection{Experimental Setup}
To test our simulator implementing the sensor node initialization method, we used
the following set-up. The sink is running on a DELL Vostro 1500 Laptop 
(1.6 GHz CPU, 2GB RAM, WinXP Pro SP2) connected with
a USB Web Camera (Microsoft LifeCam VX6000, up to 30 frames/sec, 
live video streaming of resolution $640X480$ pixels) and a wireless printer. 
The webcam can be replaced with any similar camera with a frame rate 30 fps or
higher, without any modification to the existing simulator.  The camera is set
in NON\_STOP video capturing mode and frames are taken setting the
camera in preview mode.  Camera controller is added to the simulator to allow
adjusting the focus, tilt and pan of camera as needed.

The transmitting side of the simulator runs on a DELL desktop computer (1.8 GHz
CPU, 1 GB RAM, WinXP Pro SP2) connected with LEDs on breadboard (Figure \ref{fig:real}) 
through parallel port (DB25 Connector).  
The laptop and the desktop computer are connected with our university's
wireless connection (54 Mbps).  Figure \ref{fig:sim} has a snapshot of our set-up.



\subsection{Usability Testing}
\label{sec:usability}

In order to test how our method fares with non-expert users, and especially to
figure out if the users are easily and correctly able to discard the failed
sensor nodes based on the result screen (and/or print-out), we performed a usability
study. 

\vspace{0.5mm}

\noindent \textbf{Testing Framework.} For creating an automated testing
framework, we extended the transmitter application running on the desktop
computer by implementing the usability testing and user feedback collection
functionality on it. The sink application running on the laptop was
configured to send the result (indicating passed or failed sensor nodes) to the
desktop application, as soon as it was determined. As there is no interface on
breadboard using which the users can turn off the failed sensor node(s), we
simulated the ``turning off'' mechanism in the desktop application.  As soon as
the desktop application receives the result from the laptop application, it
shows the layout of the sensor node field (i.e., the breadboard) on screen,
associating each sensor node with a transparent button with the layout of the sensor node
in the background.  The users are instructed to transfer the result from the
laptop screen to the desktop screen by clicking on the buttons (on the desktop
screen) corresponding to the failed sensor nodes shown on the laptop screen. After
test completion, the desktop application has the functionality of showing the
questionnaires to obtain user feedback and logging the data.  In our current
tests, we did not make use of the printed output.

\vspace{0.5mm}

\noindent \textbf{Test Cases.} We created five categories of test cases to
evaluate our method against different types of possible attacks and errors.
These included (1) matching SAS and no synchronization errors (to simulate
normal execution scenarios, where no attacks or faults occur), (2) (single- and
multiple-bit) SAS mismatch on a varying number of sensor nodes; (3) missing,
pre-mature and delayed turning on of the Sync LED (to simulate synchronization
errors), (4) both SAS mismatch and synchronization errors, and (5) variable
distance (from 0.5 to 2 feet) between the camera and the transmitters.  Ten
test cases for each category were created.  Each user executed a total of five
test cases, one each selected randomly from each of the five categories.

A (portion of the) screenshot of the result of execution of one of the test cases is shown in
Figure \ref{fig:res2}.

\ifthenelse{\boolean{eps}}{}
{
\begin{figure}[htbp]
\begin{center}
\fbox{\includegraphics[keepaspectratio,height=3.3in,width=2.5in]{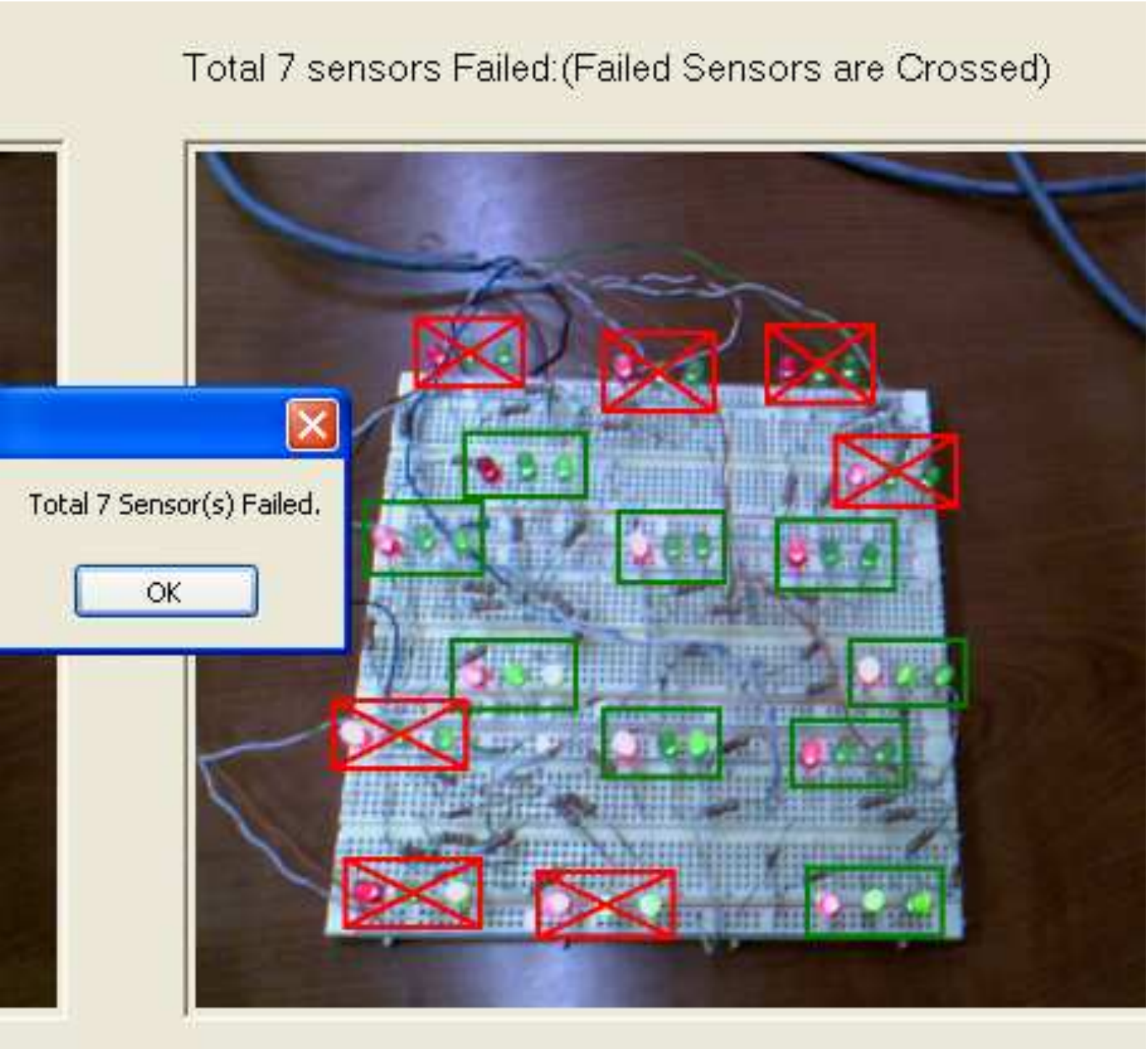}}
\caption{Result Screen: 7 Failed Sensor Nodes (marked by ``red cross''), 9 Passed Sensor Nodes (marked by ``green rectangle'')}
\label{fig:res2}
\end{center}
\end{figure}
}


\vspace{0.5mm}

\noindent \textbf{Test Participants.}
We recruited 21 subjects for our usability testing. Subjects were chosen on a
first-come first-serve basis from respondents to recruiting posters and email
ads. At the end of the tests, the participants were asked to fill out an on screen
questionnaire through which we obtained user demographics and their feedback on the method
tested.

Recruited subjects were mostly university students, both graduate and
undergraduate, with CS and non-CS backgrounds. This resulted in a fairly young
(ages between 22-31 [\textit{mean}=$25.48$, \textit{se}=$0.5417$]),
well-educated
participant group. All participants were regular computer users. 19 out of 21 participants reported they have previously used
a PC camera (for internet chat).  None of the study participants reported any
physical impairments that could have interfered with their ability to complete
given task.  The gender split was: 17 males and 4 females.

\vspace{0.5mm}

\noindent \textbf{Testing Process.}
Our study was conducted in a graduate student laboratory of our university.
Each participant was given a brief overview of our study goals and our
experimental set-up. Each participating user was then asked to follow on-screen instructions on the
laptop and desktop computer. No training of any sort was given. Basically, the
participants played the role of the administrator in the sensor node initialization
method, as depicted in Figure \ref{fig:role2}. 
Sink output, user interactions throughout the tests and timings were logged automatically
 by the testing framework.



After completing the deputed test cases in the above manner, the participants
were asked to give some qualitative feedback on how easy or hard they found to
focus the camera on all LEDs, to read the result of the output screen and about
the overall ease/difficulty of the method. Participants demographic information
such as age, gender, educational qualification, visual disability, computer and
camera experience is also collected through this questionnaire. All user data
and feedback was logged by the testing framework for future analysis.

\vspace{0.5mm}

\noindent \textbf{Test Results.} Each of our 21 subjects executed 5 test cases,
leading to a total of 105 test cases. Most of the test cases executed
successfully giving expected results. In some cases, however, we observed a few
errors, which we categorize and describe below.

\vspace{0.3mm}
\noindent {\textit{1. Camera Adjustment Error:}} We configured our
usability testing application in such manner that if all the LEDs are not
within the camera viewpoint, an error message is shown to the user asking him/her to re-execute.
In our tests, 2 users failed to adjust the camera on one occasion each and thus
they had to repeat the tests. Therefore, the rate of camera adjustment error equals
$\frac {2}{(105+2)}\times 100\%=1.87\%$ of test cases.  
 
\vspace{0.3mm}
\noindent {\textit{2. Sink Mis-reading Error:}} Sometimes the sink is not able to
correctly read the SAS string(s) transmitted by one or more sensor nodes. This could
happen when the camera is too distant ($>$ 2 feet) from the sensor nodes or due
to reflection of LED light on the table and other nearby surfaces. In our
tests, this type of error occurred for a total of 7 sensor nodes, where SAS strings of 1
or 2 sensor nodes were mis-read in some 5 test cases. In 105 tes tcases, the sink dealt
with a total of $(105\times 16)=1680$ sensor nodes on breadboard and out of them 7
sensor nodes failed due to sink errors. So, rate of sink mis-reading error equals
$\frac{7}{1680}\times 100\%=0.417\%$.  Note that all of these errors were only
false positives, i.e., the mistakenly marked a passed sensor node as a failed one.

\vspace{0.3mm}
\noindent {\textit{3. User Error:}} A user error occurs when the user is not able to
correctly transfer the result, from the laptop screen to the desktop screen
(simulating switching off of the failed sensor node).  In our tests, 3 users
accidentally clicked, on one occasion each, a passed sensor node on the desktop
screen (this implies that a passed sensor node was turned off). However, it is
important to note that on no occasions did a user miss clicking on a failed
sensor node. In other words, we did get a few false positives but no false negatives
whatsoever. Thus, rate of user errors from our tests turned out to be equal to
$\frac{3}{1680}\times 100\%=0.18\%$.

\vspace{0.3mm}

The average time taken by each user (over the 5 test cases), to complete Steps 2
to 4 of Figure \ref{fig:role2}, is depicted in Figure \ref{fig:userTiming}.  As
we see, the time taken by all of our users to perform a test is less than a
minute [\textit{mean}=$26.5$ seconds, \textit{se}=$1.37$]. Note that these
numbers arise when we assume a fairly conservative setting, one where both
normal scenarios and attacks or faults occur with equal likelihood. However, in
practice, attacks or faults are less likely. Therefore, considering only the
normal test case, we find that that on an average a user only takes $19.18$
seconds [\textit{se}=$1.11$] to complete the whole process.

\ifthenelse{\boolean{eps}}{}
{
\begin{figure}[h]
\begin{center}
\includegraphics[height=1.5in,width=3.4in]{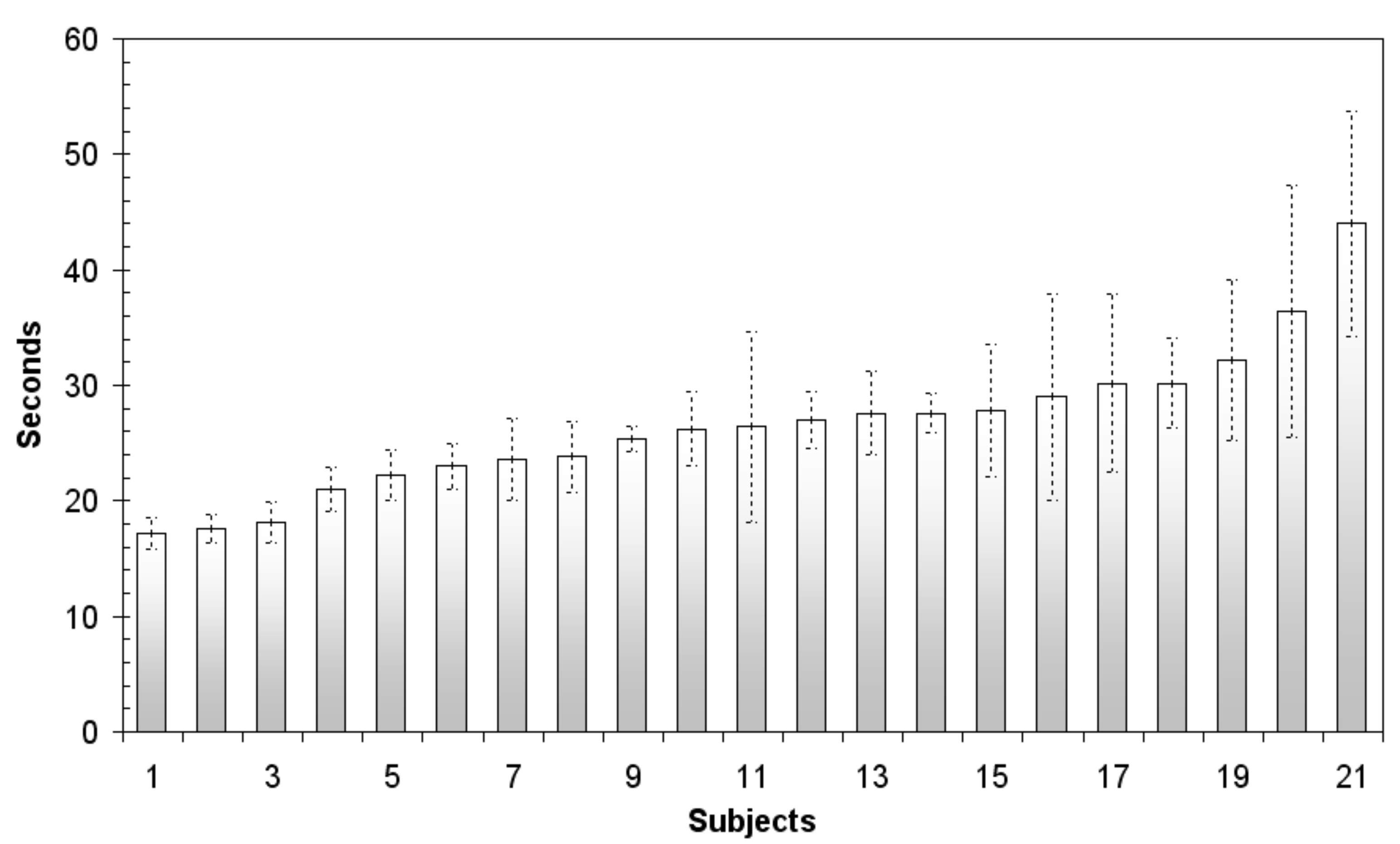}
\caption{Average time (per test case execution) taken by 21 subjects with standard error. Subjects sorted by average time.}
\label{fig:userTiming}
\end{center}
\end{figure}
}

The results we obtained through the user feedback questionnaire are shown in
Table \ref{fig:user_testing}.  Clearly, most users found the method robust and
quite easy to work with.

\begin{table*}[htbp]
\begin{center}
\centering
{\footnotesize{
{
\begin{tabular}{|c|c|c|c|c|c|c|}
\hline
  Easiness  & Very Easy & Easy & Medium Difficult & Difficult & Very Difficult & Impossible \\
\hline
       Camera Adjustment to LEDs & 5 & 13 & 3 & 0 & 0 & 0 \\
\hline
       Detection of Failed Sensor Nodes & 11 & 10 & 0 & 0 & 0 & 0 \\
\hline
       Easiness of Mechanism & 7 & 10 & 4 & 0 & 0 & 0 \\
\hline
\end{tabular}
}
}}
\caption{User Feedback (numbers denote the number of users)}
\label{fig:user_testing}
\end{center}
\end{table*}

\section{Discussion and Conclusion}
We proposed a novel method for secure initialization of sensor nodes. Based on
our testing with the method, we make the following conclusions. 


\noindent \textbf{Efficiency}: Using N Data LEDs and one Sync LED per sensor node,
the transmission requires $[\lceil \frac{20}{N}\rceil +3]\times 250$ ms. This
is equal to 3.25 sec for N=2 and 20-bit SAS data. Extraction of SAS data
from captured frames and displaying the result on screen require less than 3-4
seconds. So, execution time of the method is 7-8 seconds. Overall, as our
experiment results show, most users took less than a minute to perform the
whole process.  Also, as shown in \cite{secon}, most existing commercial sensor
motes (e.g. Mica2) can efficiently execute (within a minute) the public key
operations (private and public key generation, and one exponentiation).  Note
that these operations constitute the dominant costs in the SAS
protocol (of Figure \ref{fig:seka}) that a sensor node executes with the sink.  The
sink, on the other hand, is assumed to be a computer with a fairly strong
computational power and therefore can efficiently execute $n$ parallel protocol
instances with each of the sensor nodes.\footnote{The protocol of Figure
\ref{fig:seka} works with sink's permanent public/private keys.} 

Based on the above numbers, we recommend setting $\Delta=2$ minutes, as the
time period (to complete Steps 2 to 4 of Figure \ref{fig:role2}) by which the
key initialization will be accepted by each sensor node, by default. As our
experiments show, within 2 minutes, a human user can safely complete the
initialization process, turning off any (failed) sensor nodes, if necessary.


\noindent \textbf{Power Requirements}: From \cite{secon}, we know that most
available commercial sensor motes can do public key crypto operations using only a
small amount of power. Now, we show that the SAS data transmission through
blinking LEDs also incurs a minimal overhead on sensor motes in terms of power. For
20-bit SAS data transmission, the three LEDs on each sensor mote light-up
13 times (for a period of 250ms), i.e., for a
duration of $13\times 250$=$3.25$seconds.  Each LED has a drop voltage, $V$=
$2.9$ Volts (typical range 1.7-3.3 Volts); Current Rating, $I$= $2.2$ mA
(typical range 2-3 mA). Therefore, the maximum energy consumption per sensor mote (3 LEDs),
E=$3\times(V\times I\times t)$= $3\times(2.9 \times 2.2\times 10^{-3}\times 3.25)$ 
Volt-A-seconds =$0.062205$ Joules. As stated in \cite{secon}, the Energizer No. E91, two AA
batteries used in Mica2 motes, have a total energy of $2\times (1.5\times 2.850
\times 3600)$=$30780$ Joules.  So, our SAS data transmission requires
$\frac{0.062205}{30780}\times 100\%=0.0002\%$ of battery life of Mica2. As
shown in \cite{secon}, public key generation requires 0.816 Joules of energy.
Thus, our SAS data transmission is more that 13.11 times better than the public
key generation in terms of power consumption 


\noindent \textbf{Robustness}: Our method is quite robust to varying distances
between the transmitter and receiver. The distance between the camera and
sensor motes on breadboard can be up to 2 feet. The method also works quite well in
varying lighting and brightness conditions as it deterministically learns the
environment using the first two, All-OFF and All-ON, frames in each session.
The method could fail in presence of background noise during transmission and
reception of SAS data. It will not work in blackout or blackout in the middle
of transmission. Huge variations of lighting conditions during transmission of
SAS data which exceed color threshold of LEDs or shaking/displacement of all
sensor motes/camera while transmission of SAS data exceeding the dimension threshold
of LEDs will also cause failure of the method.  However, these will only lead
to false positives and not to an attack.  Except for the camera adjustment
errors (as discussed in Section \ref{sec:usability}), all errors occurring with
our method are localized i.e, if a single sensor mote fails due to some reason, only
that particular sensor node needs to be re-initialized. Note that this is unlike the
MiB scheme of \cite{MiB07}, where any errors lead to the
re-initialization/re-keying of the whole batch of sensor motes. Even when camera
adjustment occur in our method, only the SAS data transmission needs be repeated, not the
whole initialization process. On the other hand, MiB is less prone to user errors
than our method. However, our results indicate that our user errors only lead
to false positives and are negligible nevertheless.  In our future work, we
plan to explore how default rejection (as opposed to our current default
acceptance mechanism) would impact the efficiency, usability and scalability of
our method. It will clearly improve security.


\noindent \textbf{Scalability}: Our method can be used to to initialize
multiple sensor nodes per batch. We tested the method with 16 sensor nodes having three
LEDs each.  By using good quality wide-angle cameras (which will somewhat
increase the overall cost of the system), this number can be greatly improved,
we believe.  We are currently exploring ways to make our method more scalable.
Note that increase in the number of sensor nodes will come at only a slight cost of
increase in the length of SAS data. For example, to support 128 sensor nodes, we
would need to transmit 22 SAS bits.


\noindent \textbf{Usability}: Via a systematic usability study, we find that
our method is quite user friendly. It does not require any expertise or prior
training. Little or no acquaintance with the method is enough to administer the
process. It is easy to work with and enables easy detection of failed sensor nodes
by observing the result on the screen of the sink. Unlike the MiB scheme of
\cite{MiB07}, the administrator does not have to deal with a specialized and
often cumbersome Faraday Cage. Of course, the administrator has to deal with a
camera in our method, however, most users are getting more and more familiar
with cameras as they become ubiquitous. Moreover, a camera can be used for
purposes other than key distribution and is thus not truly specialized. Also
note that the sensor motes per batch do not need to be homogeneous. They can have
different number, color of LEDs, in any topology whatsoever (the only
requirement being they all possess one RED colored LED to act as the Sync LED).
Recall that this is unlike MiB \cite{MiB07}, which can only support homogeneous
sensor motes with very similar weights. We consider this as an important issue with
respect to usability -- an administrator might need to initialize a diverse
pool of sensor motes and should not need to group them up.


\noindent \textbf{Simplicity and Economic Viability}: The sink needs
only a camera and each sensor nodes require at least two LEDs (one Sync and one Data)
which are very cheap and commonly available. In fact, most existing commercial
sensor motes have three LEDs. Our method is quite economic, as opposed to MiB
\cite{MiB07} which requires a specialized Faraday Cage and an additional
sensor mote (called ``keying devices'') having USB interfaces.


\noindent \textbf{Resistance to Malicious Sensor Nodes}: Our method offers a natural
protection against corrupted or malicious sensor nodes\footnote{A manufacturer could
possibly sneak in malicious sensor node(s) along with normal sensor nodes shipped to a
customer, as pointed out in \cite{MiB07}.}. Our method is based on an
authenticated key exchange protocol following the security model of
\cite{CK01}. This model guarantees that an adversary who learns session key(s)
corresponding to some corrupted session(s), does not learn any information
about the keys corresponding to other uncorrupted sessions. This is unlike MiB
\cite{MiB07}, where a single corrupted sensor node can compromise keys corresponding
to all other sensor nodes\footnote{\cite{MiB07} suggests using a software-based
attestation technique \cite{SWATT} to prevent this attack.}.


\bibliographystyle{IEEEtran}   
{\footnotesize
\bibliography{SensorInit}}

\end{document}